\documentstyle[letter,preprint,epsf,wrapft]{ptptex}

\def\lsim{\ \lower-0.4ex\hbox{$<$}\kern-0.80em\lower0.8ex\hbox{$\sim$}\ }
\def\gsim{\ \lower-0.4ex\hbox{$>$}\kern-0.80em\lower0.8ex\hbox{$\sim$}\ }

\preprintnumber{
UTAP-257}
\title{
Quantum Kinetics of Deconfinement Transitions \\ in Dense
Nuclear Matter}
\subtitle{Dissipation Effects at Low Temperatures}    

\author{
Kei {\sc Iida}   
\footnote{E-mail address: 
iida@utaphp1.phys.s.u-tokyo.ac.jp} 
}

\inst{
Department of Physics, University of Tokyo, \\ 7-3-1 Hongo, Bunkyo,
Tokyo 113, Japan
}

\recdate{
\today
}

\abst{
Effects of energy dissipation on quantum nucleation of
two-flavor quark matter in dense nuclear matter encountered 
in neutron star cores are examined at low temperatures.
We find that low-energy excitations of nucleons and electrons
reduce the nucleation rate exponentially
via their collisions with the surface of a quark matter droplet.
}

\begin{document}

\maketitle

The possible existence of quark matter in neutron stars
has been studied for the past two decades from the viewpoint
of equilibrium phase diagrams (see, e.g., Refs.\ 1)$\sim$3)). 
Recently, Glendenning\cite{rf:1} relaxed the constraint 
of $local$ charge neutrality, as usually assumed, 
and discovered that a phase where quark matter and nuclear matter 
in $\beta$-equilibrium coexist in a uniform sea of electrons could
intervene between bulk quark and nuclear matter phases
for a finite range of pressures. 
This is because the presence of
strange and down quarks plays a role in reducing the electron Fermi
energy and in increasing the proton fraction of nuclear matter. 
Glendenning\cite{rf:1} and Heiselberg et al.\cite{rf:2} also claimed 
that the mixed phase could exhibit spatial structure 
such as quark matter droplets embedded in nuclear matter 
as a result of surface and Coulomb effects.

As a  star whose core consists of 
nuclear matter in $\beta$-equilibrium 
spins down or accretes matter from a companion star,
the central density could become sufficiently large for the mixed phase to 
be stable. Whether the mixed phase actually appears, however, depends on 
the occurrence of the dynamical processes leading to its appearance.
The first of these processes should be deconfinement;
the direct conversion of nuclear matter to three-flavor quark matter
that is more stable than two-flavor quark matter is inhibited, 
since relatively slow weak processes are involved.
If quark matter is deconfined from nuclear matter via a 
$first$-$order$ phase transition, a droplet of two-flavor 
quark matter could form in a metastable phase of bulk nuclear 
matter during the spin-down or accretion.
Iida and Sato\cite{rf:4} calculated the rates for the formation
of this droplet
via quantum tunneling at zero temperature 
by incorporating the electrostatic energy
into the potential for droplet formation.
It was found that a droplet formed via quantum fluctuations 
would develop into bulk matter due to electron screening effects
on the Coulomb energy of the droplet itself.
This result implies that the mixed phase is unlikely to occur.
 
The effects of energy dissipation on 
the quantum kinetics of
the deconfinement transition,
which were ignored in the previous work,\cite{rf:4} 
may have a decisive role in determining
not only
the degree of overpressure required to form an up 
and down 
quark matter droplet 
but also
the crossover temperature from the nucleation via 
quantum tunneling 
to that via thermal activation; its comparison with the temperature of
matter in neutron star cores is significant.
In this paper, we thus estimate to what extent
the rates for droplet formation via quantum tunneling
are reduced by
the friction exerted on the droplet,
according to the work of Burmistrov and Dubovskii\cite{rf:5} 
that gave useful expressions for such reduction
on the basis of a path-integral formalism developed
by Caldeira and Leggett.\cite{rf:6}
It is found to be the ballistic motion of low-lying excitations 
of nucleons and electrons that controls the dissipative processes
in the situation of interest here. 
The possible effect of neutron and proton superfluidity,
which acts to decrease the number of available quasiparticle 
states with momenta close to the Fermi surfaces,
is also considered.
Hereafter, we generally 
take $\hbar=c=k_{B}=1$.

The static properties of the deconfinement transition
are described using simple phenomenological models for 
the energies of nuclear and quark matter, as in Ref.\ 4).
For nucleons,
we write the formula for the energy density as\cite{rf:2} 
\begin{equation}
  \epsilon_{N}(n,x)=n[m_{n}+(K_{0}/18)
  (n/n_{0}-1)^{2}+S_{0}(n/n_{0})^{\gamma}(1-2x)^{2}]\ ,
\end{equation}
where $m_{n}$ is the neutron mass, $n$ is the nucleon number density, 
$n_{0}=0.16$ fm$^{-3}$ is the nuclear saturation density, $x$ is
the proton fraction, $K_{0}=250$ MeV is the
nuclear incompressibility, and $S_{0}=30$ MeV and 
$\gamma=1$ determine the symmetry term. For quarks we adopt an energy
density based on the bag model,\cite{rf:2}
\begin{equation}
  \epsilon_{Q}(n_{u},n_{d})=\left(1-\frac{2\alpha_{s}}{\pi}\right)
  \left(\sum_{q=u,d}\frac{3\mu_{q}^{4}}{4\pi^{2}}\right)+B\ ,
\end{equation}
where $q=u, d$ denote up and down quarks, 
$n_{q}$ is the number density of $q$ quarks, and
$\mu_{q}=(1-2\alpha_{s}/\pi)^{-1/3}(\pi^{2}n_{q})^{1/3}$ is the chemical
potential of $q$ quarks. 
In Eq.\ (2), all quark masses have been taken to
be zero, and the QCD fine structure constant and the bag constant
have been set as $\alpha_{s}=0.4$ and $B=120$ MeV fm$^{-3}$. 
The energy density of the electrons, which are relativistically degenerate,
is $\epsilon_{e}(n_{e})=\mu_{e}^{4}/4\pi^{2}$,
with the electron 
number density $n_{e}$ and the electron chemical potential 
$\mu_{e}=(3\pi^{2}n_{e})^{1/3}$.
We neglect the energy contributions of muons and hyperons;
the former was shown to 
be unimportant to the deconfinement transition itself,\cite{rf:4}
whereas the latter will be considered elsewhere.
Thermal effects on the static properties of the deconfinement transition
may be safely omitted; for the temperature range considered,
$T<10$ MeV, we find $T\ll \mu_{i}-m_{i}$ $\sim100$--$1000$ MeV, 
where $i=e$, $n$ (neutron), and $p$ (proton), and $\mu_{i}$ is the
chemical potential of $i$ particles
including their rest mass $m_{i}$.
The two electrically neutral bulk phases of interest
are the $\beta$-stable nuclear phase, in which $\mu_{n}=
\mu_{p}+\mu_{e}$, and the quark phase arising from the above-mentioned
nuclear matter via deconfinement at fixed pressure $P$.
The pressure $P_{0}$ of the static deconfinement transition 
was obtained\cite{rf:4} from the comparison of Gibbs free energies per
baryon between these two phases as $P_{0}\cong445$ MeV fm$^{-3}$,
where the baryon density jumps from $\sim5n_{0}$ to $\sim8n_{0}$.

The time $\tau$ required to form a single quark matter
droplet via quantum tunneling
in the initial metastable phase consisting
of nuclear matter in $\beta$-equilibrium was evaluated\cite{rf:4} 
as a function of $P$
within the WKB approximation\cite{rf:7} 
in the case in which no energy is dissipated.
Let us now obtain a formula approximately reproducing
the values of $\tau$ so evaluated. 
According to the results of Ref.\ 4), 
we expect that
the contribution of the electrostatic energy $E_{C}$
to the potential for droplet formation
may be omitted by virtue of the electron screening effects on the
droplet charge stemming from the pressure equilibrium 
between quark and nuclear matter.
We thus write the Lagrangian for 
a fluctuation of the droplet radius $R$ as\cite{rf:4}
\begin{equation}
   L=(1/2)M(R){\dot R}^{2}-U(R)\ .
\end{equation}
Here $M(R)$ is the effective mass derived from the nucleon 
kinetic energy as\cite{rf:4}
\begin{equation}
   M(R)=4\pi\epsilon_{N}(n_{\rm init},x_{\rm init})
   (1-n_{b,Q}/n_{\rm init})^{2}R^{3}\ ,
\end{equation}
where $n_{\rm init}$ and $x_{\rm init}$ are the nucleon number density
and proton fraction outside the droplet, and 
$n_{b,Q}=(n_{u}+n_{d})/3$ is the baryon density inside the
droplet, and $U(R)$ is the potential given by the minimum work 
needed to form the droplet as\cite{rf:4}
\begin{equation}
   U(R)=(4\pi R^{3}/3)n_{b,Q}(\mu_{Q}-\mu_{\rm init})
   +\Delta E_{e}(R)+4\pi\sigma R^{2}\ ,
\end{equation}
where 
$\mu_{\rm init}=[\epsilon_{N}(n_{\rm init},x_{\rm init})
+\epsilon_{e}(n_{e,{\rm init}})+P]/n_{\rm init}$ is 
the chemical potential for the bulk nuclear matter with the electron 
number density $n_{e,{\rm init}}$,
$\mu_{Q}=[\epsilon_{Q}(n_{u},n_{d})+\epsilon_{e}(n_{e,{\rm init}})
+P]/n_{b,Q}$ is the chemical potential for the bulk quark matter
embedded in the electron gas of number density $n_{e,{\rm init}}$, 
$\Delta E_{e}(R)$
is the excess of electron energy over the initial value, and
$\sigma$ is the surface tension which, being poorly known, 
is taken to be $\sim10$ MeV fm$^{-2}$.  
Within the framework of the linear
Thomas-Fermi approximation\cite{rf:8}   
in the determination of 
the electron distribution,
which proves\cite{rf:4} useful for $R\lsim 10$ fm,  we obtain
$\Delta E_{e}(R)=4\pi\rho_{Q}R^{3}\mu_{e,{\rm init}}/3e$, 
where $\rho_{Q}=e[(2n_{u}-n_{d})/3-n_{e,{\rm init}}]$
is the excess charge density of the droplet, and 
$\mu_{e,{\rm init}}$ is the initial chemical
potential of electrons.
The quark number densities $n_{u}$ and $n_{d}$
are determined independently of $R$
from the pressure equilibrium
between quark and nuclear matter and from the flavor conservation
before and after deconfinement.\cite{rf:4} 
If one takes the energy $E_{0}$ of the zeroth bound state 
around $R=0$ to be much lower than the barrier height $U_{0}$, 
an analytic formula for $\tau$ is obtained as\cite{rf:9}
\begin{equation}
\tau=\nu_{0}^{-1}\exp(A_{0})\ , 
\end{equation}
where $A_{0}$ is the doubled underbarrier action given by
\begin{equation}
   A_{0}=(5\sqrt{2}\pi^{2}/16)
         |n_{b,Q}/n_{\rm init}-1|
	 \sqrt{\sigma\epsilon_{N}(n_{\rm init},x_{\rm init})}
	 R_{c}^{7/2}\ 
\end{equation}
with the critical droplet radius
$R_{c}=3\sigma/[n_{b,Q}(\mu_{\rm init}-\mu_{Q})
-\rho_{Q}\mu_{e,{\rm init}}/e]$, satisfying $U(R_{c})=0$, and
$\nu_{0}$ is the frequency of the oscillation around $R=0$,
$\nu_{0}=[12\pi^{3/2}\Gamma(1/4)^{-2}]^{4/7}
(4\pi\sigma R_{c}^{2})^{5/7}$ $\times[M(R_{c})R_{c}^{2}]^{-2/7}$.
Naturally $\nu_{0}$ is of order $10^{23}$ sec$^{-1}$. 
As Fig.\ 1 illustrates, 
the values of $\log_{10}\tau$ given by Eq.\ (6) 
at $\sigma=10$ MeV fm$^{-2}$
agree fairly well with the computed values\cite{rf:4} 
in the pressure 
\noindent
\begin{figure} 
	        \epsfxsize=11cm   
	        \epsfysize=10cm   
	        \centerline{\epsfbox{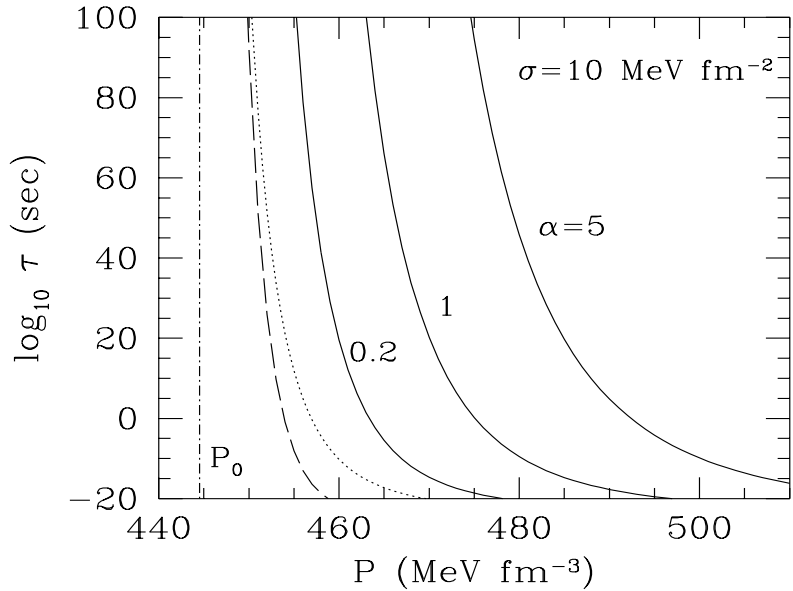}}
                \caption{
The formation time $\tau$ of a single droplet with the surface tension 
$\sigma=10$ MeV fm$^{-2}$ as a function of pressure. The dashed line
is the numerical WKB result$^{4)}$ in the absence of dissipation; 
the dotted line is the result given by Eq.\ (6); the solid lines are 
the results for dissipative cases obtained from Eq.\ (9) as 
$\tau=\nu_{0}^{-1}\exp(A)$ at $\alpha=0.2,1,5$. $P_{0}$ denotes the 
pressure of the static deconfinement transition.}
          \label{fig:1}
\end{figure}
range of interest. 
Discrepancies between these two results, which are shown to be $\sim20$,
arise mainly from the approximation 
$E_{0}\ll U_{0}=(4/27)4\pi\sigma R_{c}^{2}$
made in Eq.\ (6) and partly from
the approximation $E_{C}=0$ made in Eq.\ (5); 
the latter reduces the barrier height and width
by $\lsim10$ \%, and hence the order of the nucleation time 
by $\lsim20$. Hereafter, 
the Lagrangian (3) and the time (6) 
will thus be used as models for the formation of a 
droplet being in a completely coherent state.

Complete coherence 
can be achieved only when all the excitations 
adjust adiabatically to the motion of
the droplet surface. During the fluctuation of $R$, 
however, 
the low-energy excitations of nucleons and electrons 
appearing as quasiparticles in the vicinity of the Fermi surfaces
undergo relatively slow relaxation processes, 
leading to dissipation of the total energy of the droplet.
The fluid of quarks, which is assumed to be incompressible and hence
static,$^{4)}$ 
does not contribute to such dissipation.
Since the mean free paths of $i$ excitations ($i=n,p,e$)
are estimated as
$l_{i}\sim
n_{i}^{-1/3}(\mu_{i}-m_{i})^{2}T^{-2}>10^{2}$ fm with
the number density of $i$ particles 
($n_{e}=n_{p}\sim n_{0}$ and $n_{n}\sim4n_{0}$),
it is obvious
that the $l_{i}$ are large compared with the critical droplet radius 
$R_{c}$ ranging 1--10 fm. 
Consequently, the excitations, which behave ballistically, 
collide with the droplet surface and the resulting transfer
of momentum flux exerts an Ohmic frictional force
on the droplet given by\cite{rf:5} 
\begin{equation}
F=-16\pi\left(\frac{n_{b,Q}}{n_{\rm init}}-1\right)^{2}
\left(\sum_{i}\alpha_{i}n_{{\rm nm},i}p_{F,i}\right)R^{2}{\dot R}\ ,
\end{equation}
where 
$n_{{\rm nm},i}$ is the number density of 
a normal component of $i$ particles, 
$p_{F,i}=(3\pi^{2}n_{i})^{1/3}$ is the Fermi momentum 
of $i$ particles, 
and $\alpha_{i}$ is a factor of order unity
that depends on the properties of $i$ excitations
and on their interactions with the droplet surface.
For simplicity, we assume $\alpha_{e}=\alpha_{n}=\alpha_{p}\equiv\alpha$. 
It is instructive to note that $F\sim -M(R){\dot R}v_{F}/R$, 
where $v_{F}=(\sum_{i}n_{{\rm nm},i}p_{F,i})
/[\epsilon_{N}(n_{\rm init},x_{\rm init})+\epsilon_{e}(n_{e,{\rm
init}})]$ is the averaged Fermi velocity. 

We proceed to evaluate the effects of the  
dissipative processes mentioned above on the rates 
for droplet formation via quantum tunneling.
First, we consider a system in which 
nucleons and electrons are in a normal fluid state 
at any temperature.
The dissipation effects result in the multiplication of
the tunneling probability $\exp(-A_{0})$
by a reducing factor\cite{rf:6} $\exp(-\Delta A)$;
the expressions for $A=A_{0}+\Delta A$ were 
obtained\cite{rf:5} for an Ohmic frictional force such as Eq.\ (8) 
in the two limits  $A_{0}\ll\Delta A$ and $A_{0}\gg\Delta A$.
By noting that 
the typical velocity $\sqrt{2U_{0}/M(R_{c})}$ 
of a virtual droplet moving 
under the potential barrier 
is one order of magnitude smaller 
than $v_{F}$ $(\sim 0.4c)$,
we find that the dissipation
effects dominate the tunneling probability
in the system considered here.
The corresponding expression for $A$ reads\cite{rf:5}
\begin{equation}
  A=4\pi\alpha\left(\frac{n_{b,Q}}{n_{\rm init}}-1\right)^{2}
\left(\sum_{i}n_{{\rm nm},i}p_{F,i}\right)R_{c}^{4}s(T)\ ,
\end{equation}
where $s(T)$, 
the normalized underbarrier action composed of
the potential term coming from Eq.\ (5) 
and the nonlocal dissipation term 
controlling the dynamics of the droplet in configuration space, is 
approximately 1.3
along the extremal trajectory 
for temperatures up to the quantum-classical crossover value
$T_{0}$, as will be discussed below.
The energy dissipation may also alter the oscillation 
frequency $\nu_{0}$ by about one order of magnitude 
because of the difference
between $v_{F}$ and $|\dot{R}|$,
but this has little effect.
As depicted in Fig.\ 1,
the nucleation time $\tau=\nu_{0}^{-1}\exp(A)$ derived from Eq.\ (9) 
decreases exponentially with increase in the overpressure 
$\Delta P=P-P_{0}$, a feature resulting from the narrowed and lowered
potential barrier.

In a neutron star, the formation time of the first droplet 
is $\tau/N$, where $N$
is the number of virtual centers of droplet formation in the star. 
$N$ is roughly estimated to be 
$V/(4\pi R_{c}^{3}/3)$ $\sim 10^{48}$, where $V$ is the volume of
the central region in which the value of $\tau$ determined by
the pressure remains 
within an order of magnitude,
by calculating the pressure and density 
\noindent
\begin{figure}  
	        \epsfxsize=11cm 
	        \epsfysize=10cm 
	        \centerline{\epsfbox{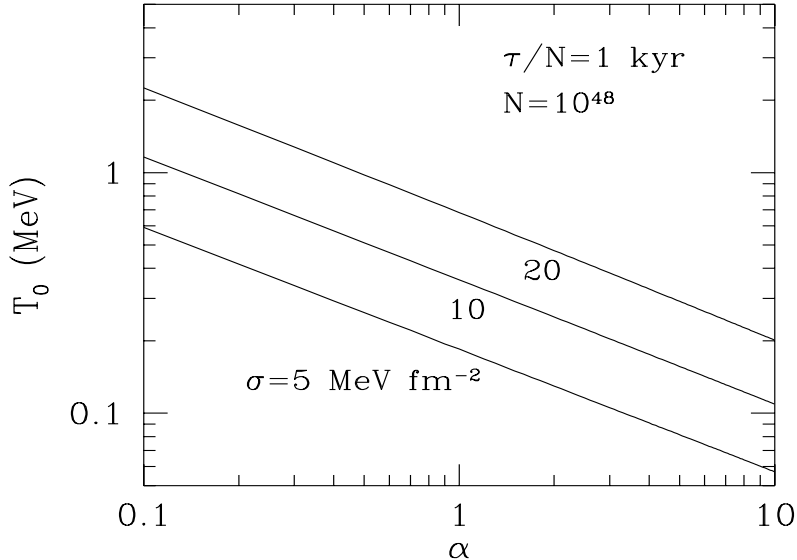}}
                \vspace{+0.0cm}
       \caption{
The crossover temperature $T_{0}$ from the quantum tunneling to 
the thermal activation regime as a function of $\alpha$ at 
$\sigma=5,10,20$ MeV fm$^{-2}$, $N=10^{48}$ and $\tau/N=1$ kyr. 
}
          \label{fig:2}
\end{figure}
profiles of the star 
from the Tolman-Oppenheimer-Volkoff equations and the equation of 
state given by Eq.\ (1). From these calculations we have confirmed that
the central pressure of the star with maximum mass, $\approx730$
MeV fm$^{-3}$, is high compared with the pressure range of interest.
In order that the first droplet may form during a time scale  
of the spin-down or accretion $\tau_{1}\sim$ yr--Gyr, 
the overpressure should amount to
such a value $\Delta P_{c}$ as to make $\tau/N$
comparable to $\tau_{1}$.
Taking as an example $\sigma=10$ MeV fm$^{-2}$, $\alpha=1$ and 
$N=10^{48}$, we see  
that the dissipative processes 
change $\Delta P_{c}$ 
from $\sim7$ MeV fm$^{-3}$ to $\sim21$ MeV fm$^{-3}$ and
$R_{c}$ from $\sim4.1$ fm to $\sim1.9$ fm.
Correspondingly, the crossover temperature $T_{0}$ from the
quantum tunneling
to the thermal activation regime,
which is derived from the relation\cite{rf:5} $A=U_{0}/T_{0}$,
goes from $\sim1.7$ MeV to $\sim0.4$ MeV. 
In Fig.\ 2 we have shown the crossover temperature $T_{0}$ 
as a function of $\alpha$ at $\sigma=5,10,20$ MeV fm$^{-2}$
and $N=10^{48}$.
This figure suggests that the first-order
deconfinement transition in nuclear matter 
of temperature $\lsim0.1$ MeV, 
as found in a neutron star core, 
if it occurs dynamically, 
is likely to proceed via quantum nucleation, although
no definite conclusion can be drawn in the absence of 
the exact information about $\alpha$ and $\sigma$.

Finally, we consider the system in which neutrons (protons) are
in a superfluid state for temperatures below the critical 
value $T_{c,n}$ $(T_{c,p})$. 
Whether or not $T_{c,n}$ and $T_{c,p}$ are finite
for densities of about $5n_{0}$ is not well known, but 
it is expected from
the properties of 
nucleon-nucleon interactions that
the $^{3}P_{2}$ neutron pairing and the $^{1}S_{0}$
proton pairing
may possibly sustain a superfluid state (e.g., Ref.\ 10)). 
Such a state plays a role in reducing the number densities
of the $n$ and $p$ excitations and thus 
in weakening the decelerating effects
of the energy dissipation on the nucleation.
In order to estimate the maximal influence of the superfluidity,
we set $n_{{\rm nm},n}=n_{{\rm nm},p}=0$ in Eq.\ (9), an assumption
valid for $T\ll T_{c,n}$ and $T\ll T_{c,p}$.
Accordingly, only
the collisions of the electron excitations with the droplet surface
contribute to the energy dissipation, leading to
$v_{F}\sim0.06c$, which is comparable
to the velocity of a droplet moving under the potential barrier. 
These considerations show that
for $\alpha\gsim1$
the nucleation time continues to be
determined mainly by the dissipative processes
rather than by the reversible droplet motion.
For the foregoing parametric combination $\sigma=10$ MeV fm$^{-2}$, 
$\alpha=1$ and $N=10^{48}$, we obtain $\Delta P_{c}\sim11$ MeV fm$^{-3}$,
$R_{c}\sim3.2$ fm and $T_{0}\sim1.0$ MeV.

In summary, we have found that the effects of the 
energy dissipation on the formation of a $u$ and $d$ quark matter
droplet via quantum tunneling in a metastable phase consisting of 
$\beta$-stable nuclear matter,
governed by the ballistic motion
of the nucleons and electrons excited 
in the vicinity of the Fermi surfaces,
act to increase the critical overpressure $\Delta P_{c}$ considerably
at low temperatures relevant to matter in neutron stars.
It is to be noted that, for better estimates of this increase,
density fluctuations 
associated with these excitations in the medium
should be taken into account, 
since the hydrodynamic and thermodynamic descriptions 
of the kinetic and potential energies 
given by Eqs.\ (4) and (5) are 
taken for granted in the present theory of
the quantum nucleation based on Ref.\ 9).
The mechanism of the growth of the nucleated droplet 
in the pressure range $P\gsim P_{0}$ also remains to be examined;
it may 
be controlled by the dissipative processes associated with 
the excitations in the hydrodynamic ($l_{i}\ll R$)
or ballistic ($l_{i}\gg R$) regime,
or possibly by the detonation arising from
the density jump due to deconfinement.
Finally, we believe that 
the present analysis can be used for the study of
the nucleation of quark matter in 
dense nuclear matter of temperatures $\sim10$ MeV,
as is of interest in stellar collapse.

\vspace{0.1cm}
The author thanks Professor K.\ Sato and Professor T.\ Satoh for
valuable comments. This work was supported in part
by the Grant-in-Aid for Scientific Research from  
the Ministry of Education, Science and Culture of Japan  
(07CE2002 and 4396).

\end{document}